# Professional Perspectives on Community and Public Engagement with Fusion Energy

**Jared Owens**

University of Michigan-Ann Arbor

**Shanna R. Daly**

University of Michigan-Ann Arbor

**Porter Malcolm**

University of Michigan-Ann Arbor

**Aditi Verma**

University of Michigan-Ann Arbor

**Acknowledgments**

The authors would like to thank the participants for their time and effort in participating in the interviews. All participants, regardless of their background, knowledge, or perspectives, demonstrated exceptional professionalism, thoughtfulness, and receptiveness to discussing the interview topics, making the project both insightful and enjoyable. The authors would also like to thank Sylvie Duffalo for her assistance in the first stage of data analysis.

**Abstract**

Fusion energy technology is rapidly approaching demonstrator deployment, but relatively little consideration has been given to publics' perceptions of or engagement with the technology. This research explores the perspectives of diverse professionals involved with the US fusion energy industry as to their knowledge on, valuation of, and preparedness for community and public (C&P) engagement on fusion energy. Participants perceived multifaceted value in C&P engagement from both industry and personal perspectives including an expressed desire for self-involvement in engagement activities, though ideally with increased support from home organizations and from engagement experts. Participants also revealed framings of fusion C&P engagement that align with deficit and contextual models of science communication thinking. These findings suggest an opportunity for collaboration between fusion professionals and engagement experts beyond one-way science communication to more participatory engagements with implications for the future of fusion energy.

## 1 Introduction

From the creation of the first controlled fusion devices in the late 1950s through today, the allure of harnessing reliable, low-carbon energy from fusion reactions has prompted research from plasma physics and electromagnetics to materials science and scientific computing (Biron, 2025; Department of Energy [DOE], n.d.; Mullen, 2025; Phillip, 1983; Government Accountability Office [GAO], 2023). Despite the lack of a guaranteed pathway to fusion's viability for commercial energy production, advancements made over decades of research have yielded insights that now drive the development of a growing fusion energy industry (DOE, 2022b; Fusion Industry Association [FIA], 2026b). Today, dozens of companies across the United States are working to develop fusion machines to bring the long-awaited technology into commercial reality, with some technologies so mature as to be in the early stages of deployment in demonstrator facilities (DOE, n.d.; FIA, 2026b; Hines-Acosta, 2025). In its most recent annual report, the FIA found that well over half of responding companies indicated a belief that, within the next 15 years, fusion energy technology will not only reach the grid, but will also reach commercial viability (FIA, 2026b).

While much of the fusion energy research to date has focused on advancing fusion's technical feasibility as an energy technology (the pursuit of consistent, net-positive energy production), meaningful attention to the full spectrum of deployment considerations will ultimately be important for reaching commercial viability. Such considerations include not just typical economic (Lindley et al., 2023; MIT Energy Initiative, 2024) and regulatory (Dorman, 2023; MIT Energy Initiative, 2024) elements, but also societal perspectives on fusion energy. Looking beyond the important but narrow permission-based objective of seeking a social license for fusion energy (Gupta et al., 2025; Hoedl, 2022; GAO, 2023), design literature in other technological development contexts points to potential benefits from incorporating not just communicatory, but participatory engagements into fusion energy's design and deployment (Kawamoto et al., 2025; Owens et al., 2024). Any such engagements, however, require careful upfront planning or what Owens et al. (2024) term "participatory design preparedness" to maximize the likelihood of achieving intended outcomes.

Through 22 semi-structured interviews, we investigated the landscape of fusion community and public (C&P) engagement knowledge, valuations, and preparedness among diverse technical and nontechnical fusion professionals, guided by the following research questions:

RQ 1. What do fusion energy professionals know about C&P engagement practices, engagement education, and public perceptions of fusion energy?

RQ 2. How do fusion professionals perceive the value of fusion C&P engagement?

RQ 3. What fusion C&P engagement variables do fusion professionals consider most important?

RQ 4. What challenges and forms of support do fusion professionals perceive as most important for fusion C&P engagement?

## 2 Background

### 2.1 The State of Fusion Energy

The drive for fusion energy comes against the backdrop of an increasing worldwide demand for reliable, low-carbon energy (IEA, 2025). Recognizing the possibility of fusion energy and describing it as "a potential game-changing clean energy technology" (Office of Science and Technology Policy [OSTP], 2022), the US government across recent federal administrations has expressed significant interest in fusion technology (DOE, 2025b) as manifested through the Department of Energy Milestones-Based Fusion Development Program established in 2022 (DOE, 2022a) and the creation of Fusion Innovation Research Engine Collaboratives in 2025 (DOE, 2025a).

Fusion energy systems utilize the same process that powers the sun: nuclear fusion reactions in which atoms combine and release energy. Controlling fusion reactions on Earth, however, has required the development of a variety of inertial and magnetic confinement technologies including tokamaks, stellarators, Z-pinch systems, laser systems, and others (GAO, 2023). At a fundamental level, the key variables involved in initiating and sustaining fusion reactions are time, temperature, and fuel density (Madsen & Miklos Gaspar, 2021), the combination or "triple product" of which is a common measure for comparing fusion energy systems.

Advancements from the earliest fusion energy systems to today have been extraordinary, with modern triple products orders of magnitude higher than early examples (Meschini et al., 2023). Despite these advancements, a number of significant technical challenges remain. The temperatures (exceeding 100,000,000 °C) and intense neutron radiation involved in fusion reactions, for example, necessitate advancements in materials science to build sufficiently durable systems (GAO, 2023). Components and materials such as advanced steels, magnets, and lithium blankets, along with fuel sourcing for deuterium and tritium, will require standing up or expanding exotic supply chains (FIA, 2026a; Madsen & Miklos Gaspar, 2021). Additionally, fusion energy systems will generate significant quantities of low-level radioactive waste that must be managed appropriately (El-Guebaly, 2023; Gonzalez De Vicente et al., 2022; GAO, 2023). Overcoming these challenges will be necessary to establish commercial fusion energy's feasibility, but demonstrating fusion's viability as an economically sound competitor to existing energy technologies will require yet further advances in technology, such as efficient and reliable systems for converting fusion reaction energy into usable electricity. As of publication, no fusion facility has produced more energy than is needed to sustain itself, let alone net-positive energy production in an amount that justifies the cost of fusion energy as a commercial energy source (GAO, 2023).

### 2.2 Public Knowledge and Valuation of Fusion

Looking beyond the technical challenges associated with fusion, research into public engagement with fusion energy has largely been limited to the past five to ten years (Oltra, Giacometti, Boso, et al., 2026; Owens et al., 2024; GAO, 2023). Even today, a 2026 scoping review of fusion social science research yielded a final corpus of only 14 publications (Oltra, Giacometti, Boso, et al., 2026). The limited body of fusion social science research that does exist has focused primarily on European countries and, to a lesser extent, the United States, reflecting both the underdeveloped state of fusion social science specifically and the wider fusion research and development gap between the Global North and Global South (Carayannis et al., 2022; Oltra, Giacometti, Boso, et al., 2026). Despite these limitations, the literature has revealed important insights into fusion energy from public perspectives. In the United States and in Europe, public awareness of fusion energy is limited but somewhat positive where present (Oltra, Giacometti, Boso, et al., 2026). At a more granular level, a number of demographic attributes have been found as statistically significant factors associated with support for fusion energy, with increasing age, political conservatism, wealth, and being male all associated with increased support (Cale & Kromer, 2015; Gupta et al., 2025; Oltra, Giacometti, Čok, et al., 2026). Beyond simple support or lack thereof for fusion, research has also shown underlying cultural contexts to play an important role in how fusion is framed by the public (Edwards et al., 2019; Prades López et al., 2008): “wider social and cultural perspectives and values play an important role in shaping the perception of fusion and other ‘future technologies’” (Prades López et al., 2008, pp. 100–101). Simply put, how people come to understand and perceive emerging technologies, including fusion, is tied to the social and cultural contexts within which individuals are situated. These individual contexts meaningfully influence how new information about emerging technologies is processed and weighed, as it is done so in relation to the hopes, concerns, and other unique attributes of each person’s unique perspective of the world.

Perhaps the most distinct theme across the relevant literature on public perspectives of fusion is the significant impact of mental associations between fusion and fission, or between fusion and the word “nuclear.” One study noted preexisting attitudes toward fission as the greatest single variable associated with support for fusion (Oltra, Giacometti, Čok, et al., 2026). Another found that increased association between fission and fusion correlated with reduced support for fusion, and that word associations with “fusion energy” tended to be more positive compared to “nuclear fusion” (Gupta et al., 2025, p. 15). Several studies speculate that this finding results from mental associations with the powerful “brand image” or “stigma” associated with existing nuclear power technologies (Gupta et al., 2025; Horlick-Jones et al., 2012; Oltra, Giacometti, Boso, et al., 2026; Oltra, Giacometti, Čok, et al., 2026). As put in one study:

> “It is as though a nuclear label serves to very powerfully communicate a rich and detailed collection of ideas and images…the response seems to be an instantaneous emotional sense of fear prompted simply by the technology being labelled by the word nuclear” (Horlick-Jones et al., 2012, p. 520).

**2.3 Engagement Rationales and Implications for Fusion Energy Technology**

Despite the fact that fusion technology is still many years away from being put to use for commercial energy production in any meaningful way, C&P engagement on fusion energy is immediately relevant. First, there are significant ethical implications associated with early C&P engagement. While contestable, it has been argued throughout the engagement literature that engagement on the direction of technology development is an important element of democratic societies (Fiorino, 1990; Kaplan et al., 2021; Secko et al., 2013; Wickson et al., 2010). Second, putting aside ethical arguments about the value of engagement, two major rationales exist for engaging with publics on fusion energy technologies: to gain useful design insights, and to build trust, both of which are particularly important early in design and deployment processes.

As for design insights, research has illustrated that, both for technology development in general and specifically for fusion, the input of non-experts can be useful given their unique perspectives (Hoedl, 2022; Prades López et al., 2008). The perspectives of non-experts allow them to foresee challenges and solutions that are often missed from experts' perspectives. Such perspectives have been leveraged in a variety of ways, including, for example, creative problem solving in scientific research tasks (Einsiedel, 2012). In line with the previously discussed findings on cultural contextual influences on framings of emerging technologies, non-experts also demonstrate awareness of social and political values and value mismatches that may be unrecognized or underappreciated from expert perspectives (Fiorino, 1990). Sometimes these value mismatches are not intrinsic to the nature of a technology itself, but rather lie in the details of its implementation. Each fusion technology comes with a unique array of not just engineering challenges, but also wider sociotechnical implications including economic, environmental, and social risks, costs, and benefits (Madsen & Miklos Gaspar, 2021; GAO, 2023). For example, fuel sourcing and safe handling is less complex for deuterium-deuterium (D–D) based fusion systems compared to deuterium-tritium (D–T), but D–D fusion involves significantly higher temperatures (and thus associated engineering challenges) compared to D–T (GAO, 2023). If and how fusion energy system designers integrate consideration of potential public value mismatches can thus directly impact downstream perspectives on the technology in deployment. Tying this to the urgency of fusion C&P engagement today, there exists an inverse relationship over time for all technologies between knowledge of a technology's actual performance and ease of modifying its design, known as the Collingridge Dilemma (Genus & Stirling, 2018). As information about a technology is assembled through design analysis, prototyping, and deployments, it simultaneously becomes increasingly difficult and cost-prohibitive to make design changes (i.e., sociotechnical lock-in (Meadowcroft, 2009)). Information about a technology's performance includes not just technical performance characteristics, but also wider sociotechnical performance including societal perspectives on the technology. It is important, therefore, to gather as much information as possible through engagements early in the design and deployment process for a technology so that course corrections, if necessary, can be made before they are prohibitively expensive and difficult to execute. This relevance of early engagement is applicable both for public engagement regarding fusion technology at large and for community engagement in specific fusion facility deployment contexts, where community engagement at the outset of a project can provide useful information for decision-making about siting and community integration (Hoedl, 2022).

As for trust building, considering both public and individual community input through engagement on fusion technology design and siting decision-making, engagements serve to build trust (Stirling, 2008; Wickson et al., 2010). As stated in a RAND report on public engagement's use in technological innovation:

> "participatory methods that include deliberation, and in particular, public dialogues and citizens' juries, have been used to measure aspects of trust in a particular technological innovation, but have also directly led to increased trust as a result of the process" (RAND Europe, 2021, p. 94).

Additionally, it is evident firsthand through the challenges associated with COVID vaccine adoption (Sapienza & Falcone, 2022) and, more recently, the siting of data centers, that failing to build trust through early and meaningful engagement can have devastating consequences on technology adoption and deployment. Given the importance of early fusion C&P engagement, it is by extension important to understand fusion industry professional perspectives of C&P engagement. Understanding C&P engagement perspectives among fusion professionals—whose own involvement in greenlighting, executing, or acting upon engagements is critical for engagement success—allows for the identification of alignments or misalignments with established findings in the engagement literature. It also helps identify areas for increasing participatory design preparedness, or "having the technical, social, and engagement-facilitation knowledge necessary to successfully execute participatory design engagements" (Owens et al., 2024).

### 2.4 Engagement Typologies

Having established rationales for community and public engagement on fusion, it is necessary to make clear the diversity of both population scopes and types of engagement that exist (Rowe & Frewer, 2005).

The appropriate population scope, type of engagement, engagement mechanism, and other planning variables can and should be drawn from the specific contextual factors surrounding an engagement, including what engagement's planners or "exercise sponsors" (Rowe & Frewer, 2005) seek to accomplish from it.

#### *2.4.1 Engagement Population Scopes*

Beginning with population scopes, the definitions of the terms adopted here are differentiated based on physical and functional proximity to fusion energy. We define **communities** as "those groups of people who live or work in close enough physical proximity to fusion energy facilities as to have a tangible and immediate interest in them" and **publics** as "those groups of people who do not necessarily have any direct connection to fusion energy facilities like communities or other stakeholders but nonetheless have interests, concerns, hopes, fears, and values that can contribute to democratic decision-making." These definitions are derived from those used in a National Academies of Sciences, Engineering, and Medicine report (National Academies of Sciences, Engineering, and Medicine, 2016, pp. 131–132) that has been cited in other fusion engagement research (Hoedl, 2022).

#### *2.4.2 Engagement Information & Power Flow*

As for the meaning of engagement, engagement can generally be defined as "seeking and facilitating the sharing and exchange of knowledge, perspectives, and preferences between or among groups who often have differences in expertise, power, and values" (National Academies of Sciences, Engineering, and Medicine, 2016, p. 131). Any particular manifestation of engagement, however, exists as one among hundreds of possible engagement formulations differentiated along "conceptually significant variables" (Rowe & Frewer, 2005) including the direction and format of information exchange, participant target population and selection methods, engagement topics, and balance of power (Bucchi & Trench, 2012; Rowe & Frewer, 2005, p. 265). These variables have led to the creation of multiple typologies to classify kinds of engagement. While some typologies are more granular than others in addressing, for example, the amount of decision-making power held by the different parties in an engagement (International Association for Public Participation [IAP2], 2018; Rowe & Frewer, 2005), a central variable is the *direction* of information flow (Rowe & Frewer, 2005). As engagement requires two or more parties, the engagement takes place between the engaged population or "engagees" (whether they are communities, stakeholders, or publics) and engagement sponsors (e.g., a fusion energy company). The baseline level of engagement, “communication,” involves one-way information flow outward from sponsors to engaged populations. This form of engagement does not involve any solicitation of information from the engaged population (Rowe & Frewer, 2005). The inverse, “consultation,” involves sponsors explicitly soliciting information from engaged populations. Sponsors may broadly acknowledge or respond to solicited information in publications (e.g., FAQ, press briefing) or even through operational changes driven by feedback, but they may do so without opportunities for active dialogue. Finally, with “participation**,”** there is at least some degree of ongoing, bidirectional information flow between a sponsor and an engaged population. To date, fusion engagement has focused largely on one-way communication of information about fusion to the public, although the fusion literature does include some discussion of participatory engagements (Kawamoto et al., 2025; Owens et al., 2024).

### 2.5 Models of Science Communication

Combining the ideas that there are multiple kinds of engagement and that the rationales for engagement exist as the foundation upon which engagements are planned, robust literature has evolved around *models* of communication specifically in the context of science and technology. The diversity of models of science communication are representative of the diversity of perspectives about what science communication means, or what it “should” mean, which is not standardized across scientific disciplines, individual scientists, or projects, with variations in thought about science communication’s primary purpose, foci, styles, audiences, messages, and channels (Kessler et al., 2022; Secko et al., 2013). One way of understanding models of science communication is as “frameworks for understanding what ‘the problem’ is, how to measure the problem, and how to address the problem” (Brossard & Lewenstein, 2009, p. 3). The explicit or implicit adoption of one or more models of science communication among scientists is not merely a matter of academic interest. Rather, it has real-world implications on if or how scientists engage with the public (Kessler et al., 2022). Although the exact formulation of science communication models varies (e.g., (Brossard & Lewenstein, 2009; Kessler et al., 2022; Metcalfe, 2022)), science communication typologies tend to segment models along similar lines to the wider engagement literature, with a break between more “communication” driven models and more “consultation” or “participation” driven models. The four models adopted here are the deficit and contextual models ("communication" type), and the lay-expert and public participation models ("consultation/participation" type) (Brossard & Lewenstein, 2009).

Beginning with the deficit and contextual models, these models are focused on largely one-way communications from science to the public. The deficit model places an emphasis on explaining and addressing underlying mistrust in science resulting from scientific illiteracy. In a fusion context, this type of thinking could manifest as, for example, media campaigns intended to raise public knowledge of fusion energy. The contextual model is similar, but adds an explicit layer of recognition to differences in identity, recognizing that the way in which people process scientific information is shaped by culture, individual circumstances, and experiences. Information delivery is still the goal of contextual model thinking, but the methods for achieving delivery are tailored to best meet the circumstances of the intended audience (Brossard & Lewenstein, 2009).

As for the more "participatory" models, these models involve a bidirectional flow of information between engagement sponsors and engagees, but they differ in their focus on why reciprocating information from the engagee is important. The lay-expertise model recognizes the relevance of expertise and knowledge possessed by the public, arguing that scientists are "often unreasonably certain—even arrogant—about their level of knowledge, failing to recognize the contingencies or additional information needed to make real-world personal or policy decisions" (Brossard & Lewenstein, 2009, p. 34). In a fusion context, this could manifest as, for example, participatory design workshops (Kawamoto et al., 2025; Owens et al., 2024). Finally, with the public participation model, the goal is still participatory engagement, but with an emphasis on the political importance of participatory engagement with science. Thinking through the participatory model entails seeking to democratize decision-making authority, to a degree, about science that is otherwise typically not in public hands.

No model is intrinsically "better" than or mutually exclusive from others. Whether in the minds of individual scientists or in any given science communication context, they tend to coexist in practice (Brossard & Lewenstein, 2009). For fusion, an awareness of the mental models of fusion professionals again allows for identification of alignments or misalignments with established findings in engagement literature, and to identify areas for increasing participatory design preparedness.

## 3 Methods

### 3.1 Research Questions

This study was guided by the following research questions:

RQ 1. What do fusion energy professionals know about C&P engagement practices, engagement education, and public perceptions of fusion energy?

RQ 2. How do fusion professionals perceive the value of fusion C&P engagement?

RQ 3. What fusion C&P engagement variables do fusion professionals consider most important?

RQ 4. What challenges and forms of support do fusion professionals perceive as most important for fusion C&P engagement?

### 3.2 Participants

The primary selection criterion for participants was professional involvement in the fusion energy industry. Participants were recruited across a range of ages, seniority in roles, prior familiarity with C&P engagements, job types, and educational backgrounds.

Recruitment took place primarily through two methods: snowball sampling, and targeted direct outreach. For snowball sampling, an initial network was formed at the 2024 Fusion Industry Association conference in Washington, DC. Each interaction included a brief introduction to the study, an invitation to participate, and an inquiry as to others who may be interested in participating. For targeted outreach, individuals working at fusion companies were identified and invited to participate via LinkedIn. Additionally, a general LinkedIn post briefly describing the study was shared to invite interested individuals to participate.

Interested applicants were directed to a Qualtrics-based consent form, pre-engagement survey, and interview scheduling tool. The pre-engagement survey collected demographic information and information about participants' education and experience related to C&P engagement. Those who completed both the pre-engagement survey and the interview were offered a $100 Visa gift card.

The demographic breakdown of participants is shown in Table 1. Ultimately, there were 22 participants from 14 organizations. Eleven of the participants came from four of the eight companies in the DOE Milestones Program (DOE, 2025a). Participant titles, education, years of experience in fusion, and roles were diverse, spanning from an undergraduate student and intern with under a year of experience in fusion to a PhD-educated company vice president with decades of experience in fusion. Approximately half of participants indicated they had received some kind of formal engagement related training, with the meaning of “engagement related training” defined as broadly as possible. Most training was minimal, such as single workshops or tangentially related academic coursework (e.g., anthropology). Most training focused on science communication, business communication, and media training.

*Table 1: Participant Demographics*

| **Demographic** | **Category** | **n** |
|---|---|---|
| Gender | Male | 15 |
| | Female | 7 |
| Race/Ethnicity | White/Caucasian | 16 |
| | African-American | 1 |
| | Asian | 3 |
| | Hispanic or Latino | 1 |
| | Prefer not to say | 1 |
| Age | 18–24 | 2 |
| | 25–34 | 7 |
| | 35–44 | 6 |
| | 45–54 | 4 |
| | 55–64 | 3 |
| Highest Edu | Some College | 1 |
| | Bachelor’s Degree | 3 |
| | Master’s or Prof. Degree | 8 |
| | Doctorate | 10 |
| Any Prior Engagement Related Training | Yes | 11 |
| | No | 11 |
| Any Prior Engagement Related Experience | Yes | 22 |
| | No | 0 |

| **Demographic** | **Category** | **n** |
|---|---|---|
| Years in Fusion | 0–5 | 15 |
| | 6–10 | 0 |
| | 11–15 | 2 |
| | 16–20 | 1 |
| | 21–25 | 2 |
| | 26+ | 2 |
| Primary Role | Technical (Engineer, Scientist, Etc.) | 11 |
| | Engagement (Public relations, Communication, Etc.) | 5 |
| | Business/Management | 6 |
| Org Type | Fusion Company (DOE Milestones) | 11 |
| | Fusion Company (Non-DOE Milestones) | 6 |
| | Fusion Lab, Collaborative Facility, or Fusion-Involved Utility | 5 |
| **Example Participant Titles** | | |
| Science & Engineering | Materials Scientist, Nuclear Engineer, Nuclear Scientist, Plasma Scientist, Physical Chemist | |
| Business and Management | Co-Founder, Chief Executive Officer, Deputy Director, Vice President of Science, Vice President of Research and Development | |
| Engagement | Head of External Affairs, Director of Engagement, Communications Manager | |

All participants indicated that they had at least some experience conducting engagement activities, although the nature and extent of those experiences varied significantly. Approximately half of participants indicated minimal experience, with the other half largely having significant amounts of experience across multiple phases of their careers in academia and/or industry. The vast majority of experiences discussed by participants were science communication related activities such as classroom engagements with students, science booths at conferences, and tours of fusion facilities.

**3.3 Data Collection**

Data were collected via semi-structured interviews. Interview protocol development was driven by the research questions and the expected breadth of participant knowledge levels and perspectives. Development took place through a multi-month series of project team meetings to refine and iterate the protocol questions. During development, the protocol was also piloted with an engagement expert whose feedback was integrated into the final version. The sections of the interview protocol, their primary relevant research questions, and sample questions are included in the supplementary materials.

Interviews were conducted on Zoom and lasted approximately one hour. Participants were first reminded of the purpose of the interview and key details from the consent form. Participants were invited to ask for clarification of questions if needed and reminded that there were no right or wrong answers. Next, participants were asked to briefly describe their work and what excited them most about it. This section was included to provide contextual information about the participant's place within the fusion industry and a transition into deeper interview questions. The first core section of the interview prompted participants to explain what "community engagement" meant to them. After providing their own explanation, participants were given a standardized working definition of community engagement and instructed to answer subsequent questions in relation to the definition. Participants were then asked to describe their personal experience with and education on engagement, along with their potential interests in future engagement training. Participants were then asked about their perception of the value, or lack thereof, in conducting community engagements in a fusion energy context. Participants were also asked about their thoughts on engagement integration in the design and deployment of future fusion energy facilities, along with potential challenges involved in doing so. Interviews closed with a brief discussion on the value of public engagement and an opportunity to share anything that they felt had not been covered.

The project was funded by the University of Michigan Rackham Graduate School and was reviewed and approved by the University of Michigan Institutional Review Board HUM#00265304.

**3.4 Data Analysis**

Following each interview, automatically generated interview transcripts were cleaned and entered into Atlas.ti qualitative analysis software. Data analysis began with a recurring sequence of independent open coding followed by team discussions in batches of 2-3 interviews for the first 13 interviews, which served to establish preliminary themes for subsequent analysis. Next, two team members (one new to the data, one who analyzed them prior) re-analyzed all 22 interviews following a similar process. Findings from this second analysis pass were further refined through discussions with the larger research team. Ultimately, each interview was analyzed in relation to each research question.

**Tables of detailed code counts and example quotations for each research question are included in the supplemental materials.**

While “community” and “public” engagement are distinct concepts in the literature that were separately introduced and defined in the interviews, many participants used the two terms interchangeably. Thus, the distinction between the two is not recognized in the findings unless otherwise specified, prompting the use of the general term “engagement.”

### 3.5 Limitations

This study has limitations related to sampling approach and scope. Participants were recruited through a combination of snowball sampling, targeted outreach, and voluntary response via LinkedIn. While these approaches enabled access to a diverse range of professionals across organizations and roles within the fusion energy sector, they may have introduced self-selection bias, particularly toward individuals with prior interest in or exposure to engagement-related activities. The sample also reflects perspectives from individuals working within organizations participating in or adjacent to major U.S. fusion initiatives, including companies in the DOE Milestones Program. As a result, the findings are most directly reflective of perspectives within this segment of the fusion ecosystem, and may not fully capture views from individuals working in less formalized, smaller, or non-participating organizations. Finally, although efforts were made to include participants with a range of experiences and seniority levels, the dataset may still reflect uneven representation across subgroups within the fusion industry. Despite this, the study provides a detailed thematic account of how fusion professionals conceptualize and engage with public engagement practices and perceptions of fusion energy.

# 4 Findings

## 4.1 RQ 1: What do fusion energy professionals know about C&P engagement practices, engagement education, and public perceptions of fusion energy?

### *4.1.1 Interest in engagement education*

The majority of participants (18 out of 22) indicated a high interest in receiving additional education on engagement, but indicated a lack of awareness of the forms and sources of engagement education. For instance, P4 indicated "I would absolutely be interested in a formal education session on how to do community engagement the right and effective way, if that exists." Further, some participants seemed to equate engagement education with the narrower topic of science communication education: "I think science communication in general is a really interesting field, so yeah, it is something that is interesting to me" (P8).

### *4.1.2 Engagement methods*

Participants indicated experience with and awareness of a variety of engagement methods. In particular, many participants spent significant time discussing their own experiences conducting classroom engagements, with audiences spanning from elementary school to graduate students. In doing so, participants highlighted the significant personal impact that such engagements can have on student engagees. In discussing this, P7 reflected on their own experience from the other side of classroom engagements as a young student during the 1960s space race: "I was in 1st grade, and I vividly remember…It set me on a path of science and technology…and it lit a fire in me that burns to this day." P9, in discussing a personal example of a facility tour with students, stated

> "...in many of these cases it seems that this is the first time these students have seen this type of science and engineering and hardware. And it just seems to make a phenomenal impact where they're actually excited, you know. You see smiling faces…and that goes really well."

Although many of the engagement methods discussed were most directly related to science communication, they were not the only ones raised. Some listening- and active-dialogue-related methods were mentioned by participants, including questionnaires (2), forums (2), open houses (1), roundtable discussions (1), and even picnics (1). In addition, participants also discussed one-on-one or small group engagements, described by P1 as "micro-interactions." Such interactions were described as important both for strengthening interpersonal relationships between fusion professionals and the public, and for their ability to “snowball” as discussed by P16: “...they're excited about it, and they want to communicate that to other people… they'll get excited, and then they'll tell their family, and then… and then it just kind of snowballs from there.”

*4.1.3 State of public knowledge according to fusion experts*

Participants commented on their perceptions of the state of public knowledge of fusion and, particularly, public conflation of fusion with fission. While participants discussed a general lack of public knowledge about fusion and the need to address it through engagement, they heavily emphasized a specific concern with public conflation of fusion and fission technologies, especially in relation to the safety, environmental, and waste profiles of the technologies. In parallel, some mentioned the challenges with using the word “nuclear” when discussing fusion due to its preexisting association with fission. As put by P21, “What do we refer to fusion as? Because if you say nuclear fusion people run to the hills. If you say fusion energy, they don’t, right?...it’s unfortunate that these word games matter, but they do.”

While recognizing the baggage associated with the word “nuclear,” not everyone supported excluding the word from conversations about fusion. As P17 noted when referring to their own company’s tendency to avoid using the word nuclear: “From the technical standpoint, it’s very hard for me to get behind that, but that’s what I have to do. That’s what they tell me to do at COMPANY, so that’s what I’ll do.” As put by P9, “brand labeling and naming only gets you so far, right, at a very base level. People are aware that you're taking atoms and fusing them. And that sure sounds like something nuclear, right. So you know, you're not going to dodge that.”

**4.2 RQ 2: How do fusion professionals perceive the value of fusion C&P engagement?**

Overall, participants placed high value on engagement. Participants identified multiple sources of value as well as a range of individuals and groups impacted by engagement activities including the fusion industry, community or public engagees, and fusion experts themselves.

*4.2.1 Value to Fusion Energy Projects and Industry*

Participants attributed significant value to engagement both for individual fusion energy projects and for the fusion industry as a whole. At a project level, raising awareness of and knowledge on fusion energy; avoiding local complications; receiving unexpected but useful, actionable feedback; and developing or recruiting a workforce were recurring discussion points. As for workforce development, participants mentioned both the recruitment of individuals in the near future and a longer-term push for fusion career pipelines. The value of receiving unexpected feedback from community members, especially to enhance engagement methods, was highlighted. P4, for example, noted a personal experience with a company and community roundtable discussion:

> "They asked questions we weren't anticipating. They were concerned about things that we weren't, [that] we didn't think they would be concerned about, and they were less concerned about things we thought would be bigger concerns. So I think it was a really good opportunity for us to kind of turn the dial of where we think the community concerns are, where we think that we have done a good job of engaging people, and where we think that we are missing." (P4)

On the industry-scale value of engagement, many of the same points were raised including workforce development, raising awareness and education, and reducing pushback or securing social license. In the context of site selection, multiple participants noted an ideal arrangement in which communities are reaching out to companies hoping to be a selected site rather than companies coming to communities and trying to convince them (or even override them) to host a facility:

> "…By basically supporting people, asking you to come to their community to build something because they get the benefit of high paying jobs and clean and safe electricity. We get the benefit of deploying our product to market. It's a win-win scenario." (P12)

At the same time, some participants did note one perceived, potential downside of engagement: slowing down development or deployment. As put by P1, "But also, if you're not careful. Then you can get into a situation where there's, you know, too many cooks in the kitchen kind of thing." When these concerns were raised, however, they were usually done so with the caveat that the concern isn't with the concept of engagement, but rather how it is executed, such as P14 stating "It requires some thought and planning, and it should be properly timed."

#### *4.2.2 Value to Engagees*

In addition to project or industry-level value, participants also described the value of engagement to those who directly participate in it. Beginning with engagees, participants perceived an *intrinsic* value of engagement for engagees including everything from general statements like “you want to be a good neighbor” (P5) to more specific examples like equitable energy transitions:

> “I think we’re pretty good at coming up with new technologies and innovating and engineering. I think we’re not so good at then implementing those technologies and not leaving others behind when we do so…when we make these transitions from one energy source to another, how that affects communities is very real.” (P1)

Another lesser mentioned but recurring point of discussion was giving people hope through engagement. P6, for example, stated “I think the value of public outreach right now is that it gives people hope about climate change and our energy future.”

*4.2.3 Value to Fusion Experts*

Several participants also stated that engagement was personally rewarding. P18, for example, described hosting a science fair engagement that they found meaningful:

> “People always ask, like, oh, you'll be back next year? You know, I'll bring my neighbor, and stuff like that. So that is one very good thing I like, you know, it's a…you know, on sad, frustrating research days, that gives me some hope and motivation that that's good, you know, there's something good.”

Other participants noted utility in conducting engagements. P10 cited a significant personal change in research direction in part based on an engagement activity they facilitated:

> “One that really sticks in my mind is this person I was giving a tour…somebody says, why does it look like that? And I said I really should know the answer to that question. And so I went and looked it up, and I actually changed fields more or less. I mean you know there were lots of other things but…because somebody asked me a question that was really, you know, it changed my mind.”

### 4.3 RQ 3: What fusion C&P engagement variables do fusion professionals consider most important?

Participants identified many important qualities of engagement activities’ nature, structure, and context. Participants raised a variety of topics they believe are worth engaging with others on, including fusion science, workforce development, community concerns, environmental impact, fusion safety, site selection, and waste management. The focus of discussing these topics was frequently framed as a one-way flow of information from the sponsor to the engagee. Across engagement methods, contexts, and topics, several engagement variables were noted. The most frequently mentioned were context dependence, listening, trust, honesty and transparency, and timing.

Many participants noted that engagements should be tailored to the people for whom they are held. P5 stated: “it would be different for different communities. But you want to ask people how would you like to be engaged in the operation stage.” To this end, participants also mentioned the importance of tailoring information shared with others to their existing knowledge and education level, with many participants noting this as their greatest personal engagement strength. As put by P12,

> “I would say that I'm pretty decent at taking really pretty complicated physics aspects and translating it and forming it into a form that can be digested by whoever the audience is, whether it's anywhere from 3rd grade up to 30 years old…and it's not like translated in a deprecating sense, it's like recognizing that people have different backgrounds and respecting those backgrounds.”

The timing of engagement was noted as important. Emphasis was placed on conducting engagement early and often, both in the context of individual facilities and of the current and projected state of the fusion industry in the not-so-distant future:

> “if you don't place importance on public communications at this early formative stage, then, when you're ready to build 100 or 1,000 or 10,000 fusion power plants. Your struggle is going to be a lot greater, and it's going to be on you…And so in that regard, I think public engagement is incredibly important, even from a field that is still, you know, highly technical” (P9).

That said, a few participants believed it possible to engage too early, such as P21, when referring to early engagement, stating: “One has to, of course, make a plan for everything, but, you know, that's a bit of a cart before the horse.”

In conjunction with timing, another widely discussed point was the interrelated concepts of trust, transparency, and honesty in engagements. Trust, transparency, and honesty extended beyond the safety aspects of fusion technology to include economics. P2 described this in a community setting, highlighting the problem of project cancellations:

> “If we start something and we get people’s hopes high, and then we pull back, it’s not good…we should be really thoughtful and intentional about the messaging and any time it looks like we’re making a commitment we should be very careful that we don’t hurt people’s trust with the utilities.”

Finally, listening was included as an important aspect of engagement by many participants, such as P16 noting that engagements should not simply be a one-way flow of information from scientists and engineers to members of the public:

> “I think that a huge part of community engagement is listening, and not just [to] immediately start blabbing about fusion without having an understanding of what they even know, right? Like, I want to make sure that I hear their side first, like, any skepticism they might have, or, like, that way it informs me of how I should communicate to them.”

### 4.4 RQ 4: What challenges and support do fusion professionals perceive as most important for fusion C&P engagement?

When it comes to challenges associated with engagement, participants noted financial and expertise-based barriers to conducting engagement activities. As indicated by some participants, with fusion companies operating on limited funds trying to conduct advanced technology research, engagement funding can be deprioritized. One participant lamented,

> “They don’t want to put out the financial resources…but if you don’t do it or you don’t do it well, you’ll be paying that and then some back when things go wrong…then you’re dumping money later to try to do the work you should have been doing, and could have been doing for a lot cheaper and better on the front end” (P5).

Engagement funding was also noted by one participant as an exacerbated issue due to the unexpected cancellation of expected government funding for engagement: “the government said it was going to help us do it so we made proposals. And that was a big part of our plan, because they said they were going to help us. That’s just dead. We’re still gonna do that stuff…but of course that limits the scope right?” (P10).

Finally, as concerns beneficial preparations and resources for engagement preparedness, participants stated that, in addition to the financial element, additional contextual information about specific engagements and assistance with identifying appropriate engagement methods would be useful. Participants mentioned that insights from talking to community leaders and conducting historical research would be helpful in tailoring engagement activities:

> "…specific concerns that communities have that I just might not have engaged with as much or might have not engaged with at all…there's often a lot of historical context that you're not always 100% up to speed on. And so there might be, like sources of animosity or, you know, concern or fear…It would be nice to brief the full context of the conversation you're about to enter" (P12).

As for methods, participants noted the importance of expanding their toolbox for conducting and evaluating engagement activities. P13 noted: "I mean, I think we could use help from, you know, people who have expertise in this…We shouldn't reinvent the wheel. I mean, things like this have been done before. And so we should use the expertise of people who know how to do these kinds of community engagement processes." As to engagement assessment methods, P14 noted in the context of one of their own experiences with a company-sponsored engagement event: "Communications would be like oh great job, but I have no idea what the audience thinks…I don't have feedback, so I don't know."

## 5 Discussion

Overall, participants demonstrated a high knowledge of, value for, and interest in C&P engagement for fusion energy. Fusion professionals' high valuation of engagement alone is an encouraging sign for the future of C&P fusion engagement, but beyond this valuation, participants demonstrated personal interest in participating in engagement activities, knowledge of engagement methods and planning variables in line with the engagement literature, a self-awareness of areas for improvement or support that can further support C&P fusion engagement, and an openness to learn more from the engagement literature and engagement experts. Our interpretation of participant responses across these findings, however, points to an implicit emphasis on deficit and contextual model thinking about the purpose of fusion C&P engagement, though curiously also supported with flashes of insight that align with lay-expert and participatory model thinking. This is not a criticism of participants, as there is a historically strong foundation of deficit model thinking among scientists and science communication experts (Brossard & Lewenstein, 2009; Secko et al., 2013). Further, there is a need for the kinds of communicatory engagements that arise from deficit and contextual model thinking, as in practice, complex engagement projects involve dimensions of all science communication models (Brossard & Lewenstein, 2009). These findings do, however, represent a potential gap in the engagement thinking among fusion professionals that could be holding back more participatory-style engagements that could potentially benefit all parties if appropriately applied. To expand on this interpretation of participant responses, the findings are mapped to models of science communication in Table 2.

### 5.1 Models of Science Communication Thinking Across Participants

*Table 2: Models of Science Communication Thinking Among Participants*

| Science Communication Models | RQ 1 Knowledge of Public Perspectives | RQ 1 Mentioned Methods | RQ 2 Value of Engagement | RQ 3 Engagement Topics | RQ 3 Engagement Variables |
|---|---|---|---|---|---|
| **Deficit** | Low knowledge of fusion<br>Conflation of fusion & fission<br>Problematic association with word “nuclear” | Classroom engagements<br>Micro-interactions<br>Event booths<br>News/social media<br>Facility tours<br>Town halls<br>Demonstrations | Raising population knowledge of fusion (general)<br>Differentiating fusion from fission<br>Differentiating fusion from “nuclear”<br>Social licensing/buy-in<br>Avoiding local complications<br>Strengthening local connections | Fusion science<br>Fusion safety<br>Fusion environmental impact<br>Fusion waste management | Context dependence (knowledge)<br>Honesty/ transparency<br>Trust/trustworthiness<br>Credibility |
| **Contextual** | Low knowledge of fusion<br>Conflation of fusion & fission<br>Problematic association with word “nuclear” | Classroom engagements<br>Micro-interactions<br>Event booths<br>News/social media<br>Facility tours<br>Town halls<br>Demonstrations | Raising population knowledge of fusion (general)<br>Differentiating fusion from fission<br>Differentiating fusion from “nuclear”<br>Social licensing/buy-in<br>Avoiding local complications<br>Strengthening local connections<br>Community/public pride | Jobs and workforce development<br>Community priorities and concerns<br>— | Meeting people where they’re at<br>Context dependence<br>Making it fun<br>Honesty/ transparency<br>Trust/trustworthiness<br>Credibility |
| **Lay-Expert** | — | Town halls | Unexpected useful feedback | Unexpected useful feedback<br>Site selection | Listening |
| **Public Participation** | — | Town halls | Right thing to do<br>Giving hope<br>Community/public pride | Community priorities and concerns<br>Site selection | Listening<br>Honesty/transparency<br>Trust/trustworthiness<br>Credibility |

Table 2 maps relevant findings from RQs 1-3 to the four models of science communication (Brossard & Lewenstein, 2009). Not every finding mapped to a specific model; for example, the engagement variable for “early timing” is universally applicable. Some findings also mapped to multiple models but in different ways; for example, participants discussed trust/trustworthiness in the context of securing a social license for fusion, while others talked about it as something intrinsically “right to do.” In the former, trust maps more directly to deficit or contextual model thinking, whereas in the latter, it maps more to public participation thinking.

At a high level, findings across RQs 1-3 largely map most directly to deficit and contextual model thinking. Both of these models emphasize a need to increase public knowledge on a scientific topic and thus favor one-way communication methods, but they differ in the focus on tailoring delivered information with an eye toward straight knowledge deficits (deficit model) vs diverse cultural contexts (contextual model) (Brossard & Lewenstein, 2009). The following quote from P8 most directly illustrates this thinking through the idea that fear of a technology is mitigated by gaining knowledge about it:

> “I think the most important thing in my mind is to mitigate fear…I think community engagement in my mind is communicating what is happening, what we know, and what we would like to find out in a manner that mitigates fear.”

Further to the focus on deficit and contextual model thinking, a major recurring theme across interviews was participants’ highlighting of perceived widespread limitations of public knowledge on fusion. These perspectives on public knowledge of fusion, including general awareness, fusion’s relation to fission, and the word “nuclear,” are all in line with what is found in the limited existing literature on the topic (Gupta et al., 2025; Horlick-Jones et al., 2012; Oltra, Giacometti, Boso, et al., 2026; Prades López et al., 2008). More specifically, much of what participants stated aligned with contextual model thinking and its focus on one-way information delivery informed by cultural considerations, particularly when participants went out of their way to emphasize the need to “meet people where they are at” and recognize the unique historical and cultural contexts of individual communities or engagement participants. The engagement methods discussed and practiced by participants also overwhelmingly fell in the category of primarily one-way communication methods. This is in line with existing literature on how scientists' mental models of science communication align with their choice of engagement methods (Kessler et al., 2022).

Despite the general focus on deficit and contextual model thinking, participants still exhibited lay-expert and public participation model thinking, such as P10's noting a research direction change after a question from a member of the public, or P5's comment on showing respect to host communities.

While the science communication literature in recent years has included a push for thinking beyond the deficit and contextual models, the literature acknowledges the utility of communicatory engagements and that multiple models and their associated methods often co-exist in practice (Brossard & Lewenstein, 2009; Metcalfe, 2022; Secko et al., 2013). Participants' general focus on deficit and contextual model thinking is therefore not wrong, but, based on the engagement literature, it is also not necessarily the most effective focus given many participants' stated engagement goals of building trust and establishing buy-in.

### 5.2 Power in Siting

Participants recognized multiple aspects of community consent and decision-making power in the development of future fusion facilities, particularly in the siting process, but through divergent lenses. From one perspective, the respect for community consent and power came from a fear of its ability to inhibit, delay, or otherwise shut down projects, necessitating the need for buy-in. This perspective again largely aligns with deficit and contextual model thinking, wherein the underlying premise is addressing pushback from the public on advancing a technology arising from a perceived lack of knowledge or understanding (Brossard & Lewenstein, 2009). It is also in line with some of the existing fusion-specific research on the value of engagement for establishing a social license to deploy the technology (Hoedl, 2022). From another perspective, the respect for community power came from its intrinsic value in a democratic society, which directly aligns with public-participation model thinking, which as a model is built on an underlying premise of the importance of public participation in decision-making about the direction of science and technology in society (Brossard & Lewenstein, 2009).

### 5.3 Expert Self-Expressed Interest in Learning About, Conducting or Participating in Engagement

Participants largely expressed personal interests in learning about and participating in fusion engagements themselves. In doing so, they frequently noted the personal, intrinsic value they gained from participating in engagements. This personal interest and intrinsic valuation of engagement ties to the participants' own perceptions of the importance of the societal impact of their work and the ethics of careers in science and engineering. It could also explain some of the hesitancy expressed by a few participants in distancing fusion from the word nuclear in that they felt it was, to a degree, misleading to do so. The finding of high self-interest in engagement aligns directly with research on scientists' likelihood to participate in engagement activities. As put in Besley et al. (2013, p. 971):

> "...a deficit model view that a lack of public knowledge is harmful, a personal commitment to the public good, and feelings of personal efficacy and professional obligation are among the strongest predictors of seeing outreach as important and in participating in engagement activities"

Given the mismatch between participants' engagement goals and underlying communication models—and participants' openness to learning more about engagement—there is a clear need for more engagement education opportunities for fusion professionals in alignment with existing literature on the need for scientists to learn more about engagement to practice it effectively (Stofer et al., 2023).

### 5.4 Implications

The results from this study indicate that there are individuals in the growing fusion sector with significant subject-matter expertise who are interested in participating in engagement activities, but face both knowledge based and institutional challenges to that participation. Given the importance of expert participation in engagement activities, supporting technical experts with financial resources, educational resources, and career-related incentives for conducting or otherwise participating in engagement activities is not only potentially beneficial to fusion companies, but also something that could serve as a work experience enhancement for those who are interested. Further, in the same way that basic science research collaboration between fusion companies is a mutually beneficial endeavor, enhanced co-development and deployment of both community and public engagement mechanisms between companies has the potential to benefit both the public and the fusion industry at large.

The research conducted on these topics by experts in academia can and should be leveraged by industry. Doing so would be serving a public good and has the potential to further, powerfully demonstrate the value of social, design, and engagement science research in the development and deployment of emerging technologies. Traditionally, fusion research collaborations between academia and industry have been heavily focused on the natural sciences and engineering involved with fusion energy technologies, but our findings suggest an opportunity for stronger collaborations between academic researchers in the social sciences and fusion experts, as envisioned in Figure 1. Figure 1 shows two ends of the spectrum of possible community and public engagements on fusion. The low end is a one-way open loop of information from fusion technical science to communities and publics with an assumption of low public knowledge, involving only the basics of science communication from the wider engagement science landscape. The high end, on the other hand, draws on elements from each science communication model to leverage knowledge from fusion, social, engagement, and design sciences. The high end integrates community/public histories and cultures as well as community/public knowledge and values in engagement planning for enhanced participatory design preparedness. Knowledge and values are also then integrated in the execution of fusion design, deployment, and scaling decision-making through participatory and co-design expertise from design science. Such an integration leverages the knowledge and expertise of participatory engagement experts, social scientists, and communities/publics toward the ultimate goal of improved design and deployment outcomes for all involved parties.

*Figure 1: Low vs. High Knowledge and Value Integration*

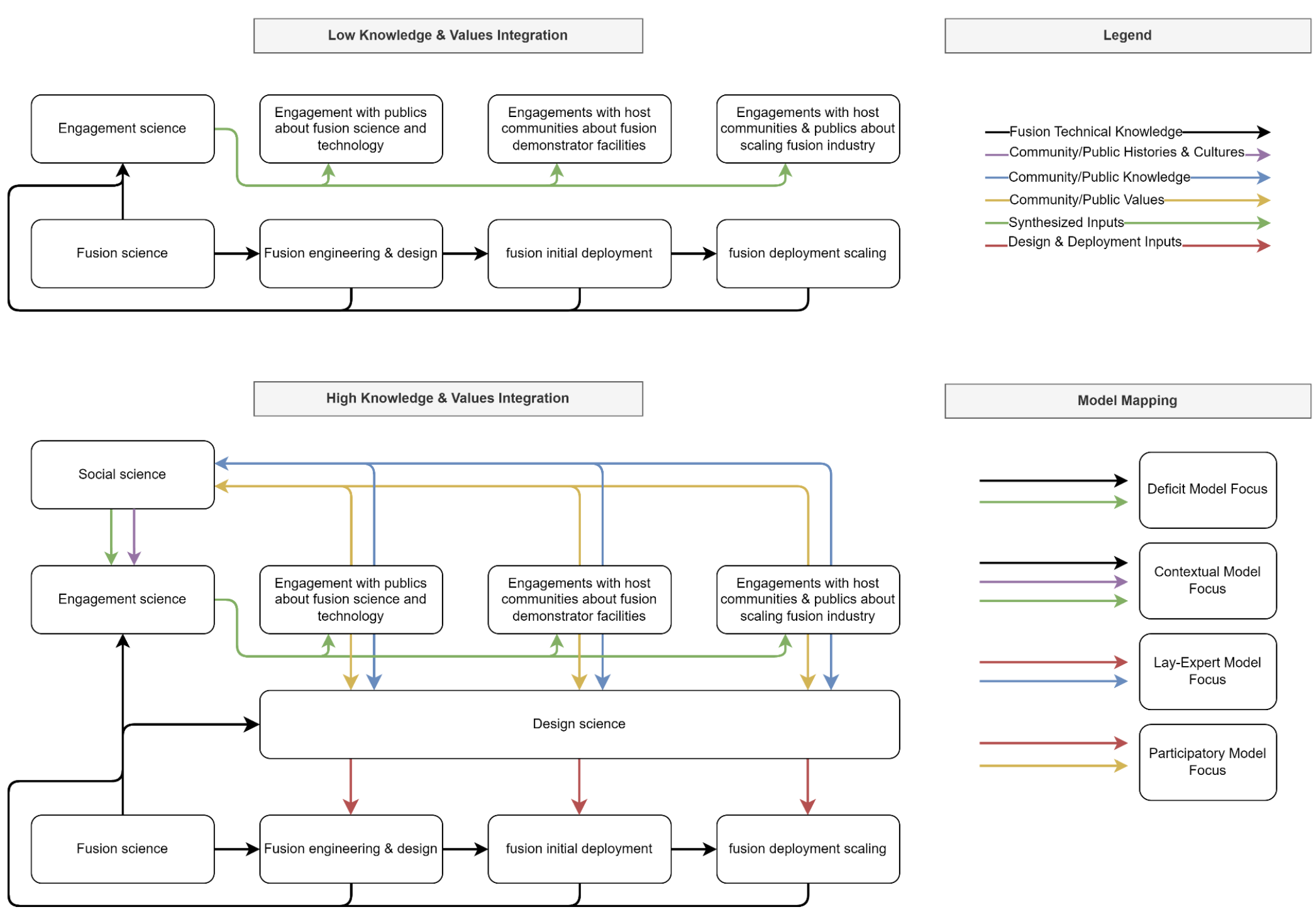

## 6 Conclusion

The deployment of fusion energy technology, long seen as an ultimate energy source of either fiction or the distant future, is within sight. Major technological challenges remain, but the technology is now in its early stages of initial commercial demonstration. If fusion energy is to ever become a viable player in the energy sector, however, it will involve not just addressing the physical dimensions of fusion engineering, but the social and cultural dimensions as well. Consideration of the social side of fusion and its impact on lives is not just morally relevant, but critical to establishing a sociotechnically sustainable environment for fusion's deployment and scaling. This interview based study explored the perspectives of diverse professionals in the fusion energy industry as to their knowledge of, perspectives on, and preparedness for community and public engagement on fusion energy technologies. The findings indicate both exceptional promise and potential pitfalls in future fusion engagements. Fusion professionals are eager to participate in engagement and recognize its importance, but they often have a self-acknowledged limited understanding of the full landscape of engagement types and mechanisms, and lack institutional structures or support for participating in engagement if it is not their primary job function. These findings reveal a significant opportunity for diverse experts' collaboration on the sociotechnical dimensions of fusion energy to enable fusion's future in a way that benefits communities, publics, and companies alike.

## Supplemental Materials

*Supplemental Table A: Engagement Population Scopes in a Fusion Energy Facility Context*

| Scope | Working Definition in a Fusion Energy Facility Context | Examples in Fusion Context |
|---|---|---|
| Communities | Those groups of people who live or work in close enough physical proximity to fusion energy facilities as to have a tangible and immediate interest in them. | Fusion facility workers, local residents, local businesses |
| Stakeholders | Those groups of people with personal or professional interests in a fusion energy facility, but do not necessarily live or work in close proximity to it. | Fusion facility investors, regulators, supply chain businesses |
| Publics | Those groups of people who do not necessarily have any direct connection to fusion energy facilities like communities or other stakeholders but nonetheless have interests, concerns, hopes, fears, and values that can contribute to democratic decision-making. | State residents, national residents |

*Supplemental Table B: Models of Science Communication*

| Model | Direction of Information Flow | Underlying Primary Value | Focus |
|---|---|---|---|
| Deficit Model | One-way communication (Experts -> Communities/Publics) | Building trust, securing “social license” | Focus on addressing knowledge deficits |
| Contextual Model | One-way communication (Experts -> Communities/Publics) | Building trust, securing “social license” | Focus on addressing knowledge deficits *in a way that is informed by cultural context* |
| Lay-Expert Model | Bidirectional, participation (Experts <-> Communities/Publics) | Recognizing the intrinsic value of other forms of knowledge | Focus on acknowledging the value of knowledge from outside the traditional boundaries of western science |
| Participatory Model | Bidirectional, participation (Experts <-> Communities/Publics) | Recognizing the intrinsic value of empowering non-experts in democratic societies | Focus on empowering others in decision-making about emerging technologies |

*Supplemental Table C: RQ 1 Discussed Engagement Methods*

| Method | Example | Count |
|---|---|---|
| Classroom engagements | …go talk to the kids in the school system. They're gonna go talk to their parents. (P2) | 7 |
| Micro-engagements | I've had conversations with people on airplanes you know, just like, random people. (P13) | 7 |
| Event booth | …there was a <large event> in <city> where we had a booth, and I was a part of that and helped, you know. Got to talk to a lot of different people… (P1) | 7 |
| News/media coverage | …but what you really have to do is make sure you take full advantage of the initial, if you will, bandwagon the initial excitement right? NIF actually achieved ignition, you know they have a 60 Minutes special. They have all this coverage. (P9) | 7 |
| Facility tours | …what we do here at <company> is that sometimes we give tours of the facility to the locals or people just visiting. (P14) | 6 |
| Town halls | …just various town hall events around the area for communicating about what fusion is and what we're working on. (P1) | 6 |
| Demonstrations | …so I would volunteer to go to different schools and do science experiments with them. Like, bring a plasma ball, or bring a fluorescent light and explain there's plasma in there, that kind of stuff to middle school, elementary, or high school kids. (P17) | 6 |
| Questionnaires | What are people concerned about when people learn about this? What might be their probable concerns, that sort of thing. And so it was kind of a tall order, and we did a couple of things. One is we sourced a questionnaire kind of a tool, a survey tool…(P5) | 2 |
| Forums | You probably want to have an open forum, where you talk about what you know. Be honest, about what risks exist, about what's going on, also what opportunities exist, and what benefits will exist for the community. (P10) | 2 |
| Open houses | You're not gonna have like this, you know, every month, you know, we're having an open house for questions. because you've answered everybody's question. (P2) | 1 |
| Roundtable discussions | We held what we think is a very thoughtful fusion energy roundtable discussion at our offices. We thought very long and hard about what community groups we wanted involved in that discussion, and we invited a handful of organizations and individuals from those organizations to our offices to have a 2 h discussion about what is fusion? How does it impact you based on your role in the community? (P4) | 1 |
| Picnic | I mean not to say that you wouldn't want to have community engagement, but it might look like an annual, you know, picnic at the site for families or something like that (P2). | 1 |

*Supplemental Table D: RQ 1 Participant Impressions of Public Perspectives on Fusion*

| Impression of Public Perspectives | Example | Count |
|---|---|---|
| Conflation of fusion and fission | ...community engagement has to be where we start, for that is a place where we can differentiate to the public that fusion is not fission, while they overlap in some areas and ways. But we have to differentiate that because there's certain, like, the stigmas that come along with fission, right, that we don't have with fusion. So it's just basically informing the public, informing the community that look, there are certain safer aspects of fusion versus fission… (P3) | 12 |
| Little public knowledge of fusion | You know if I walked out here and asked people in my community about fusion, maybe 1 out of 10 would be able to explain what it is, a state of readiness, that sort of thing. So there's your audience (P5) | 8 |
| Problematic association with word nuclear | I think the most obvious to me is the sort of taboo that comes around the word nuclear…there is a pretty intentional effort, I think, in most fusion companies, to distance ourselves from fission and not use those words. And to me that's all motivated by community engagement, and like how the public perceives the work that we do (P1) | 6 |

*Supplemental Table E: RQ 2 Perceived Value of Engagement*

| Value Recipient | Engagement Value | Description | Example | Count |
|---|---|---|---|---|
| Fusion project or industry at large (positive) | Differentiating fusion from fission | Ensuring public understanding of the differences between fusion and fission | Breaking the ties between fusion and fission, and letting people know…even if it's not impacting them right now (P6) | 12 |
| | Social licensing/buy in | Establishing community or wider public buy-in | The primary goal is to get them to accept your existence (P10) | 11 |
| | Jobs/Workforce development | Building a fusion workforce | Not thinking about what the workforce of the future is going to look like…completely ignoring all of the human effort that is going into this massive project (P4) | 10 |
| | Raising population knowledge of fusion (general) | Generally increasing knowledge of fusion | One of those that we've been really tackling is public education, engaging the communities and educating them on the role of nuclear fusion (P11) | 7 |
| | Avoiding local complications | Reducing community/local resistance or pushback on an individual project | What a waste of time this all would have been if we get to that point [deployment] and nobody wants this thing. What a failure to launch that would be (P4) | 6 |
| | Unexpected useful feedback | Unique, useful insights from non-experts | All it takes is someone asking one question to be like oh that sparks another idea, and like, maybe it can make the design better (P16) | 5 |
| | Strengthening local connections | Building a positive relationship between a company/facility and its site community | We want them [people in the community] to be on our side (P13) | 3 |
| Fusion project or industry at large (Negative) | Slowing down development or deployment process | Articulated potential concern with engagement as a damper on speedy tech development or facility deployment | I could see it happening where you're kind of maybe getting so much feedback from questions from the community that it kind of bogs you down a little bit, maybe distracts you from actually doing the thing you're supposed to be doing (P16) | 4 |
| Engagee (Positive) | Giving hope | Giving people a sense of hope from promising technology | So, I think the value really is giving people hope…being able to communicate that to someone who might be struggling or doesn't have hope that things will get better…(P16) | 4 |
| | Right thing to do | N/A | You want to be a good neighbor. You don't want to site your facility in a community that doesn't want you there, basically…it shows respect to a host community (P5) | 4 |
| | Community/public pride | Building a sense of pride around the fusion industry/a fusion facility | You want them to be proud that they are living next door to a 50 or 100 MW fusion power plant (P7) | 2 |
| Fusion Expert (Positive) | Personally enjoyable/meaningful | Articulated personally meaningful element of conducting or participating in engagement | I remember, like. kids that you wouldn't expect to be like, oh, wait, this is cool, and they'd be like, they would call somebody over and be like, wait, did you see how this works? This is actually kind of cool. Like, I think that was really cool to see. (P17) | 7 |

*Supplemental Table F: RQ 3 Engagement Topics*

| Engagement Topic | Example | Count |
|---|---|---|
| Fusion safety | It's very important to make them aware of the risks as well as the safety associated with the fusion plant…(P22) | 13 |
| Jobs and workforce development | I mean, I think just like the whole. There's the relational piece, you know, like some of the folks they were asking questions and they were making connections right? Some of these people are interested in jobs working in fusion (P2) | 10 |
| Environmental impact | They need to understand that it will be safe and it will be maximum energy with a minimum amount of environmental impact (P7) | 10 |
| Waste management | At that time it may be important to talk about what is being disposed of, where it is going, and why it is being handled the way that it is handled (P8) | 8 |
| Fusion science | Just basically going to broad events with a huge stage, huge audience…And kind of telling people about this sort of science (P3) | 7 |
| Site selection | There's dialogue..and you begin to have a conversation in which there's an exchange of information that is as transparent as can be, so that they can make a decision (P5) | 7 |
| Community priorities and concerns | …what are the issues that matter to these people?...So I think there's like a genuine way to look at it, too. and just like trying to understand where these people are coming from and their concerns. (P1) | 6 |

*Supplemental Table G: RQ 3 Engagement Variables*

| Variable | Description | Example | Count |
|---|---|---|---|
| Timing: Earlier is better | Stated importance of early engagement | I think it's critical, you know, before you decide to select a site (P2) | 15 |
| Listening | Listening as part of engagement (not just speaking) | So, it was really nice to see folks in the room engaging with the material in a way that wasn't just a one-way flow of information…more of a conversation (P4) | 12 |
| Meeting people where they're at | Recognition and responsiveness to differing engagee education, knowledge, and perspectives | I try to think from their perspective what needs to be explained and then try to focus on those [things] (P22) | 11 |
| Honesty/ Transparency | Openness about fusion pros and cons | I think the open communication as you operate such a thing, as to how's it going…All that stuff should be communicated with a lot of transparency (P21) | 10 |
| Systematicity | Establishing an engagement plan, engagement as a procedure or systematic process | I think I care a lot about a just energy transition. And that's gonna look a little different from town to town. I think trying to get the local community on board is one of the most important parts. And I wouldn't personally know the best way to do that from a systematic perspective… Systematically, I'm not exactly sure how,I would have no idea how to start that. (P1) | 10 |
| Context dependance | Recognition and responsiveness to contextual differences between engagements | The way that you describe things and the priorities in which you talk about various issues depends on who you're talking to (P9) | 7 |
| Trust/ trustworthiness | Engagement with a sense of mutual trust or understanding | I think there's just been the mindset of do the thing first and then communicate it later, but I think it should be the opposite…that's how you build trust right? (P16) | 4 |
| Credibility | Established credibility | You know I have the science background. I have the street cred from the science perspective to try to, you know, engage with them (P13) | 4 |
| Making it fun | Making engagement engaging | Make sure…that there is something fun (P14) | 4 |

*Supplemental Table H: RQ 4 Barriers, Useful Preparations, and Assistance*

| Barriers or Preparation/ Assistance | Kind of Barrier or Preparation/ Assistance | Description | Example Segment | Count |
|---|---|---|---|---|
| Barriers | Company financial resources | Limited financial resources companies are willing to allocate to engagement | …A larger budget would be nice, because you can make it more fun, more engaging. (P14) | 8 |
| | Company incentives | Lack of incentive or support structure within companies to do engagement | …those activities require dedicated resources, and I can 100% tell you that in my current job that is exceedingly limited…where does that time come from, right? Where do the resources to do that actually happen? (P20) | 4 |
| | Cultural/mindset differences | Friction from personal or wider cultural attitude toward engagement | it's the attitude which we come with. You know, if we come with a very hefty attitude that I am the scientist, I know everything, I think then the interactions will not go well, so we have to be humble on our end (P18) | 3 |
| Useful Preparations or Assistance | Understanding indiv. community | Understanding the unique history and context of an individual community | Understanding the audience is a major gap that I have, and I would need to fill before I ever stood in front of someone and tried to be helpful (P15) | 11 |
| | Engagement methodologies | Building awareness of, selecting, or facilitating appropriate engagement methods | And I wouldn't personally know the best way to do that from a systematic perspective. I would have no idea how to start that (P1) | 7 |
| | Engagement consultants | Internal or external individuals/orgs for engagement consulting | Given present circumstances, I think more interactions with other people who have experience in scientific communication (P8) | 3 |
| | Engagement assessment methodologies | Building awareness of, selecting, or facilitating engagement *assessment* methods | …I think a lot of people who try to do a community engagement, something they're bad about is defining what constitutes success. (P20) | 2 |

## Interview Protocol

| Section | Sample Questions | RQs |
|---|---|---|
| Background | What excites you most about your work? | |
| Terminology | In the context of fusion energy technologies, what does “community engagement” mean to you?<br>● [Probe] What does “community” mean to you in “community engagement?”<br>● [Probe] What does “engagement” mean to you in “community engagement?” | RQ 1 |
| Community Engagement Knowledge | Have you received any training or education in community engagement or related topics?<br>● [If yes] Could you tell me a little bit about it?<br>● [If no] If given the opportunity, is training or education in community engagement something that you’d have any interest in pursuing? | RQ 1 |
| Community Engagement Valuation | In your mind, what do you see as the value of community engagement in a fusion energy facility design context? | RQ 2 |
| Community Engagement Procedures | From your perspective, looking across the entire design process for a fusion energy facility, do you think there are any stages of the process that could particularly benefit from community engagement?<br>● [If yes] At what stages?<br>○ What benefits do you foresee community engagement potentially bringing at these stages?<br>● [If no] Pass<br><br>From your perspective, looking across the entire design process for a fusion energy facility, do you think there are any stages of the process that could be particularly hindered by community engagement?<br>● [If yes] At what stages?<br>○ What issues do you foresee community engagement potentially causing at these stages?<br>● [If no] Pass | RQ 3 |
| Community Engagement Readiness | Let’s assume for the sake of discussion that you and your company will be expected to conduct or otherwise participate in community engagement on fusion energy facility design in the future. As you look to this future, do you believe that you personally have the preparation you need to effectively participate in community engagement on fusion energy facilities design?<br>● [If yes] What do you think is your greatest strength when it comes to community engagement?<br>● [If no] What preparation do you think you might benefit the most from?<br>○ What barriers might make it hard to engage with communities on design?<br>○ What support might facilitate you in engaging with communities on design?<br><br>When it comes to community engagement, what do you think is your company’s strongest internal or external resource?<br><br>When it comes to community engagement, what internal or external resources do you think your company would benefit most from? | RQ 4 |
| Public Engagement Valuation | When it comes to the design of fusion energy systems, what do you see as the value of public engagement, if anything?<br>● [If yes] Where in the design process of fusion energy systems would public engagement be useful, if anywhere?<br>● [If no] Why not? | RQ 3<br>RQ 4 |
| Close Out | Is there anything else that you wanted to share that we haven’t covered today? | ALL |